\documentclass[journal, draftclsnofoot, onecolumn, 12pt]{IEEEtran}
\usepackage{ifpdf}
\usepackage{cite}
\usepackage{amsmath,amstext,amsgen,amsfonts,amssymb}
\usepackage{enumerate}
\usepackage{algorithmic}
\usepackage{array}
\usepackage{mdwmath}
\usepackage{mdwtab}
\usepackage{epsfig}
\usepackage[tight,footnotesize]{subfigure}
\usepackage{url}
\newtheorem{theorem}{Theorem}[section]
\newtheorem{corollary}[theorem]{Corollary}
\newtheorem{lemma}[theorem]{Lemma}
\newtheorem{definition}[theorem]{Definition}

\newtheorem{proposition}[theorem]{Proposition}
\begin{document}
\title{On Inverses for Quadratic Permutation Polynomials over Integer Rings}
%
         %
         %

\author{Jonghoon Ryu \\
          Telecommunication R\&D center, \\
          Samsung Electronics, Co., Ltd., Korea. \\
          jonghoon.ryu@samsung.com \\
         \vspace{1em}
       Oscar Y. Takeshita \\
  Silvus Technologies, Inc., Los Angeles, CA. \\
         oscar@silvuscom.com

         %
         %
         \vspace{1em}
         Submitted as a Correspondence to the IEEE Transactions on Information Theory \\
         Submitted : January 23, 2011
       }
\maketitle
\markboth{Jan. 23, 2011}{Jan. 23, 2011}
\begin{abstract}
       Quadratic permutation polynomial interleavers over integer rings have recently received
       attention in practical turbo coding systems from deep space applications to mobile communications.
       In this correspondence, a necessary and sufficient condition that determines
       the least degree inverse of a quadratic permutation polynomial is proven. 
       Moreover, an algorithm is provided to explicitly compute the inverse polynomials.
\end{abstract}
\pagebreak
\section{Introduction}
Interleavers for turbo codes have been extensively 
investigated~\cite{Berrou, Sun, Takeshita:AG}.
Today the focus on interleaver constructions is not only 
for good error correction performance of the corresponding turbo codes 
but also for their hardware efficiency with respect to power consumption and speed. \\
The work in~\cite{Sun} opened the door to a class of polynomial based interleavers. 
In particular, quadratic permutation polynomials (QPP) were emphasized 
because of their simple construction and analysis. 
Their performance was shown to be excellent~\cite{Sun, Takeshita:AG}. 
The practical suitability of QPP interleavers has been considered in
a deep space application~\cite{JPL:QPP} and in 3GPP long term evolution (LTE)~\cite{LTE:standard}. \\
The inverse function for a QPP is also a permutation polynomial (PP)
but is not necessarily a QPP~\cite{JPL:QPP}. 
However, there exists a simple criterion for a QPP to admit a QPP inverse~\cite{Ryu:inverse_QPP}. 
A simple rule for finding good QPPs has been suggested in~\cite{Takeshita:AG}.
Some examples in~\cite{Takeshita:AG} do not have QPP inverses. 
Most of QPP interleavers proposed in 3GPP LTE~\cite{LTE:standard} admit a quadratic inverse
with the exception of 35 of them. \\
In~\cite{Suvitie}, a necessary and sufficient condition that determines the least degree inverse of a QPP
by using Chinese remainder theorem and presenting the inverse function as a power series is given. 
As an example, an exact formula that determines the degree of the inverse PP is shown when the degree is no larger than $3$.  \\
In this correspondence, we provide a
necessary and sufficient condition  by using linear congruence approach in~\cite[pp. 24-40]{Ryu:thesis}
that determines the degree of the inverse when the degree is no larger than $50$. 
The condition is characterized by an exact formula and consists of simple arithmetic comparisons. 
We further provide an algorithm to explicitly find the inverse PP(s).
The algorithm is suitable for implementation since it consists of solving linear congruences. \\
This correspondence is organized as follows. 
In section II, we briefly review PPs~\cite{Hardy, Rivest, Lee:null_poly, Lee:PP, Duffee} 
over the integer ring $\mathbb{Z}_N$ and relevant results. 
The main result is derived in section III, and
examples are given in section IV. Finally, conclusions are discussed in section V. 
\section{Permutation Polynomial over Integer Rings}
\noindent In this section, we revisit the relevant facts about
PPs and other additional results in number theory 
to make this paper self-contained. 
\noindent Given an integer $N \geq 2$, a polynomial $f(x) = \sum_{k=1}^{K} f_{k} x^k \pmod{N}$, 
where $f_1, f_2, \ldots,f_K$ are non-negative integers and $K \geq 1$, 
is said to be a PP over $\mathbb{Z}_N$ 
when $f(x)$ permutes $\{0, 1, 2, \ldots, N-1 \}$~\cite{Rivest, Lee:null_poly, Lee:PP}.
It is immediate that we can use this constant-free PP 
without losing generality in our quest for an inverse PP 
by the Lemma 2.1 in~\cite{Ryu:inverse_QPP}.\\
In this correspondence, let the set of primes be $\mathcal{P} = \{2, 3,  5 , ...   \}$. 
Then an integer $N$ can be factored as $N = \prod\limits_{p \in \mathcal{P}} p^{n_{N,p}}$, 
where $p$'s are distinct primes. In addition, $n_{N,p} \geq 1$, for a finite number of $p$ and $n_{N,p}=0$ otherwise. 
%
%
%
\begin{theorem}[\cite{Sun,Ryu:inverse_QPP}]
\label{thm:qpp_cond}
   Let $N = \prod\limits_{p \in \mathcal{P}}  p^{n_{N,p}}$ and 
   denote $\alpha$ divides $\beta$ over $\mathbb{Z}$  by $\alpha \vert \beta$. 
   The necessary and sufficient condition for a quadratic polynomial $f(x)=f_1 x+ f_2 x^2 \pmod{N}$ 
   to be a PP can be divided into two cases. 
   \begin{enumerate}
      \item $2|N$ and $4\nmid N$ (i.e., $n_{N,2}=1$)\\
            $f_1+f_2$ is odd, $\gcd(f_1,\frac{N}{2}) = 1$ and 
            $f_2 = \prod\limits_{p \in \mathcal{P}}  p^{n_{f,p}}, n_{f,p} \geq 1$, $\forall p$ 
            such that $p \neq 2$ and $n_{N,p} \geq 1$.
      \item Either $2 \nmid N$ or $4|N$ (i.e., $n_{N,2} \neq 1$)\\
            $\gcd(f_1, N)=1$ and
            $f_2 = \prod\limits_{p \in \mathcal{P}}  p^{n_{f,p}}, n_{f,p} \geq 1$, $\forall p$
            such that  $n_{N,p} \geq 1$.
   \end{enumerate}
\end{theorem}
\begin{theorem}[\cite{Hardy}]
\label{thm:linear_congruence} 
      Let $\alpha$, $\beta$ be any integers and $N$ be a positive integer. 
      The linear congruence $\alpha x \equiv \beta \pmod{N}$ has at least one solution 
      if and only if $\gamma|\beta$, where $\gamma = \gcd(\alpha,N)$.
      If $\gamma | \beta$, then it has  $\gamma$ mutually incongruent solutions. 
      Let $x_0$ be one solution, then the set of the solutions is 
      $$x_0, x_0+\frac{N}{\gamma}, x_0+\frac{2N}{\gamma}, \dots, x_0+\frac{(\gamma-1)N}{\gamma}$$, 
      where $x_0$ is the unique solution of 
      $\frac{\alpha}{\gamma}x \equiv \frac{\beta}{\gamma} \pmod{\frac{N}{\gamma}}$. 
\end{theorem}
%
%
%
%
%
%
%
%
%
%
\begin{definition}[\cite{Lee:null_poly, Lee:PP}]
\label{def:equivalent_poly} 
      Two  polynomials  $f_{1}(x)= \sum_{k=1}^{K} f_{1,k} x^k$  and
      $f_2(x)= \sum_{k=1}^{K} f_{2,k} x^k $ of degree $K$ are called 
      congruent polynomials modulo $N$ if $f_{1,k} \equiv f_{2,k} \pmod{N}$, where $1 \leq k \leq K$
      and equivalent polynomials modulo $N$ if $f_{1}(x) \equiv f_{2}(x) \pmod{N}$, where $0 \leq x \leq N-1$. 
\end{definition}
\begin{definition}[\cite{Takeshita:AG, Lee:null_poly, Lee:PP}]
\label{def:zero_poly} 
      A polynomial $z(x) = \sum_{k=1}^{K} z_{k} x^k \pmod{N}$ is called a non-trivial zero polynomial of degree $K$ modulo $N$
      if  $z_K \not\equiv 0 $ and $z(x) \equiv 0 $, $0 \leq x \leq N-1$.
      Specifically,  $z(x)=0 $  is a trivial zero polynomial.
\end{definition}
%
%
%
%
%
%
%
%
\begin{proposition}[\cite{Lee:null_poly, Lee:PP}]
\label{pro:zero_poly} 
      If two polynomials $f_{1}(x)$ and $f_{2}(x)$  are equivalent but not congruent,
      there exists a non-trivial null polynomial $z(x)$ such that $f_{1}(x)-f_{2}(x) \equiv z(x) \pmod{N}$.
\end{proposition}
%
%
%
%
%
\begin{definition}
\label{def:least_deg_poly} 
       Let $f(x)$ be a PP.   A PP of least degree has a 
      least degree among all equivalent polynomials of $f(x)$. 
\end{definition}
The following proposition was proposed in~\cite{Lee:null_poly, Lee:PP}. The proof is
shown for its simplicity. 
\begin{proposition}[\cite{Lee:null_poly,Lee:PP}]
\label{pro:less_than_N_poly_deg} 
      Let $f(x) = \sum_{k=1}^{K} f_k x^k \pmod{N}$, where $f_K \not\equiv 0$ and $K \geq N$.
      Then there exists an equivalent polynomial of $f(x)$ such that the degree of the equivalent polynomial is less than $N$.
\end{proposition}
\begin{IEEEproof}
      Let $z(x) = f_K \cdot x^{K-N} \cdot \prod_{k=0}^{N-1}(x-k) $. 
      Clearly $z(x)$ is a zero polynomial. 
      Let $\bar{f}(x) = f(x) - z(x)$, 
      then  $\bar{f}(x)\equiv f(x)$ but $\deg\{\bar{f}(x)\} < \deg \{f(x)\}$. 
      By applying this repeatedly,  an equivalent polynomial of degree equal to $N-1$ can be found.
\end{IEEEproof}   
\begin{proposition}[\cite{Ryu:thesis}]
\label{pro:existence_of_inverse}
      Let  $f(x) = f_1x + f_2x^2$ be a QPP and let $k$ be an integer such that $k \geq 1$. 
      Let us take $f(x)$ such that $2 \nmid f_1$ when  $2|N$ and $4 \nmid N$. 
      Then $f_1 + k f_2$ is an unit for all $k \geq 1$, i.e.,  $f_1 + k f_2$ is invertible and
      $\frac{1}{f_1 + k f_2}$ is well defined. 
\end{proposition}
\begin{IEEEproof}
      By Theorem~\ref{thm:linear_congruence}, 
      an element $f_1 + k f_2$ in integer rings $\mathbb{Z}_N$ is an unit if and only if $\gcd(f_1 + k f_2,N) =1$.  
      We show that  $\gcd(f_1 + k f_2, N) = 1$.  
      \begin{enumerate}
           \item $2|N$ and $4\nmid N$ (i.e., $n_{N,2}=1$)\\
                 In this case there exist two equivalent QPPs~\cite{Zhao}, 
                 i.e., $f_1x + f_2x^2$ and $(f_1+ \frac{N}{2} )x + (f_2+\frac{N}{2})x^2 $, 
                 where $2 \nmid f_1$.
                 Let us take a polynomial $f(x) = f_1x + f_2x^2$ such that $2 \nmid f_1$. 
                 Suppose that $\gcd(f_1 + k f_2, N) \neq 1$. 
                 Then there exists a prime $p$ such that $p | ( f_1 + k f_2)$ and $p |N$. 
                 By Theorem~\ref{thm:qpp_cond}, if $p|N$, then $p|f_2$ but $p \nmid f_1$. A contradiction.       
           \item Either $2 \nmid N$ or $4 |N$ (i.e., $n_{N,2}\neq 1$)\\
                 In this case, there exist one (if $2 \nmid N$) or two (if $4 | N$) equivalent QPPs~\cite{Zhao}.
                 In either case, by Theorem~\ref{thm:qpp_cond} and a similar argument in (1), $\gcd(f_1 + k f_2, N) = 1$.  
       \end{enumerate}
\end{IEEEproof}
Since the inverse of only one of the equivalent polynomials is sufficient for our purposes,  
$f(x) = f_1x + f_2x^2$ such that $2 \nmid f_1$ will be considered in the rest of the correspondence. 
The following corollary is an extension of Proposition~\ref{pro:existence_of_inverse}. 
\begin{corollary}
\label{cor:existence_of_inverse}
      Let  $f_1$, $f_2$ and $N$ be the integers  in
      Theorem~\ref{thm:qpp_cond} and let $k$, $k_1$ and $k_2$ be integers such that $1 \leq k_1 \leq k_2$.  
      Let us take $f(x)$ such that $2 \nmid f_1$ when  $2|N$ and $4 \nmid N$. 
      Then $\gcd\{ \prod_{k=k_1}^{k_2} (f_1 + kf_2), N \}$ is an unit.
\end{corollary}
\begin{IEEEproof}
   This is a direct consequence of Proposition~\ref{pro:existence_of_inverse}.
\end{IEEEproof}
\section{Inverses of Quadratic Permutation Polynomials}
\noindent In this section, we derive a necessary and sufficient
condition for a QPP to admit a least degree inverse 
in Theorem~\ref{thm:main_theorem} (main Theorem). We also explicitly find the inverses in Algorithm~\ref{tab:algorithm}. \\
This section is organized as follows. 
We first show that the problem of finding inverse PP(s) of least degree is equivalent 
to solve a system of linear congruences. 
Then we show that the inverses can be found by factoring the matrix for a system of linear congruences 
and solving it. 
We also show that solving the system of linear congruences can be much simplified and finally, by showing the number and the form of zero polynomials, 
we find all the inverses of a QPP. 
%
%
%
\begin{lemma}
\label{lemma:matrix_equilvalence}
      Let $f(x) = f_1x +f_2 x^2 \pmod{N}$ be a QPP. 
      Then there exists at least one inverse $g(x)$.
      Further, finding all inverse PP(s) up to degree $N-1$  is equivalent to solving a system of linear congruences,
      \begin{eqnarray}
            \mathbf{A} \mathbf{g} \equiv \mathbf{b} \pmod{N}  , \nonumber
      \end{eqnarray}
      where 
   $$ a_{i,j} = (if_1+i^2f_2)^j,      1 \le i,j \le N-1, \nonumber  $$
   $$ \mathbf{g} = [g_1, g_2, ... ,g_{N-1}]^T,  \;\;\;\mbox{and}\;\;\; \mathbf{b} =  [b_1,b_2,...,b_{N-1}]^T = [1,2,...,N-1]^T. \nonumber $$
\end{lemma}
\begin{IEEEproof}
      Since the set of PPs forms a group under function composition, 
      the existence of an inverse for a QPP is guaranteed~\cite{JPL:QPP,D_and_F}.
      Let $g(x)$ be an inverse PP of $f(x)$ and suppose that $\deg\{g(x)\} \ge N$. 
      Then by Proposition~\ref{pro:less_than_N_poly_deg}, 
      it can be reduced to an equivalent polynomial of degree less than $N$. \\
      Since $g(x)$ is an inverse, $(g \circ f)(x) \equiv x $, where $0 \leq x \leq N-1$.
      The equivalence of $(g \circ f)(x) \equiv x $ and 
      $\mathbf{A} \mathbf{g}\equiv \mathbf{b}$ is shown by evaluating
      $(g \circ f)(x) =  \sum_{k=1}^{N-1} g_k (f_1 x + f_2 x^2)^k \equiv x $
      at each point $1 \leq x \leq N-1$.
      Note that $(g \circ f)(0) \equiv 0 $ trivially holds. 
      Consequently, solving $\mathbf{A} \mathbf{g} \equiv \mathbf{b} $ is
      equivalent to finding all the inverse PP(s) up to degree $N-1$. 
      Since the number of inverse PP(s) up to degree $N-1$ is finite, 
      there exists a least degree inverse. \\
\end{IEEEproof}
%
%
%

\begin{lemma}
\label{lemma:LDU_like_decomposition}
      Let $\mathbf{A}$ be an $N-1$ by $N-1$ matrix in Lemma~\ref{lemma:matrix_equilvalence}.     
      Then $\mathbf{A}$ = $\mathbf{L} \mathbf{D} \mathbf{U}$, where $\mathbf{L}$, $\mathbf{D}$ and $\mathbf{U}$ are $N-1$ by $N-1$ matrices as shown below. \\
      $\mathbf{L}$ is an $N-1$ by $N-1$ lower triangular matrix such that  
      \[ l_{i,j} = \left\{ \begin{array}
                 {r@{\quad  \quad}l}
                 \binom{i}{j} \cdot \prod_{k=i}^{i+j-1}(f_1 + k f_2)     &   \mbox{if  } i \ge  j   \\
                  0   &  \mbox{otherwise}
                 \end{array} \right.   \; . \]  
      $\mathbf{D}$ is an $N-1$ by $N-1$ diagonal matrix such that $d_{i,i} = i!$, where $1 \le i \le N-1$. \\
      $\mathbf{U}$ is an $N-1$ by $N-1$ upper triangular matrix such that
      \[ u_{i,j} = \left\{ \begin{array}
                 {r@{\quad \quad}l}
                  1     &  \mbox{if  }  i = j  \\
                \mathbf{q}^{(i,j)} \mathbf{V}^{(i,j)} \mathbf{r}^{(j)}   & \mbox{if  }  i < j \\
                  0  &  \mbox{otherwise  } \\
                 \end{array} \right.   \; . \]  
      $\mathbf{q}^{(i,j)}$ is an $1$ by $j$ matrix such that $ q^{(i,j)}_{k} =  (i f_1+i^2 f_2)^{k-1}$, where $1 \leq k \leq j$. \\ 
       $\mathbf{V}^{(i,j)}$ is a $j$ by $j$ upper triangular matrix such that
      \[ \mathbf{V}^{(i,j)} = \left\{ \begin{array}
                 {r@{\quad \quad}l}
                 \mathbf{I}     &    \mbox{if  } i = 1  \\
            \prod_{k=i-1}^{1} \mathbf{W}^{(k,j)}   &  \mbox{otherwise  }
                 \end{array} \right.   \;  \]  
      and $\mathbf{r}^{(j)} = [0, 0,...,0, 1 ]^T $ is a $j$ by $1$ matrix.  \\
      $\mathbf{W}^{(k,j)} $ is a $j$ by $j$ upper triangular matrix such that 
      \[  w^{(k,j)}_{m,n} = \left\{ \begin{array}
              {r@{\quad \quad}l}
               0     &    \mbox{if  } m \ge n   \\
         (k f_1 + k^2 f_2)^{n-m-1}   &  \mbox{otherwise}
              \end{array} \right.   \; ,  \]
      where $1 \leq m,n \leq j$.
\end{lemma}
\begin{IEEEproof}
  See Appendix A. 
\end{IEEEproof}
The factorization in Lemma~\ref{lemma:LDU_like_decomposition} is similar to 
$\mathbf{L} \mathbf{D} \mathbf{U}$ decomposition except that $\mathbf{L}$ has not $1$s on the diagonal~\cite{Strang}. 

\begin{lemma}[\cite{Butson, Lazebnik}]
\label{lemma:LDU_like_matrix_solution}
      Let $\mathbf{A}$, $\mathbf{L}$, $\mathbf{D}$ and $\mathbf{U}$ be the matrices in Lemma~\ref{lemma:LDU_like_decomposition}.
      Then  $\mathbf{A} \mathbf{g} \equiv \mathbf{b} \Leftrightarrow \mathbf{D} \mathbf{h} \equiv \mathbf{e} $, where $\mathbf{h} \equiv \mathbf{U} \mathbf{g}$ and
 $\mathbf{e} \equiv \mathbf{L}^{-1} \mathbf{b}$. \\
  Let us identify 
$N-1$ by $1$ matrices $\mathbf{g} = [g_1, g_2,...,g_{N-1}]^T$, $\mathbf{h} = [h_1, h_2,...,h_{N-1}]^T$ with  $g(x) = \sum_{k=1}^{N-1} g_k x^k $, $h(x) = \sum_{k=1}^{N-1} h_k x^k $, respectively. Then the degree and the number of $\mathbf{g}$ and  $\mathbf{h}$
      are equal. 
\end{lemma}      
\begin{IEEEproof}
    Since all the diagonal elements of $\mathbf{L}$ are units by Corollary~\ref{cor:existence_of_inverse},
     $\mathbf{L}$ is an unit~\cite{Duffee}.
     Thus $\mathbf{A} \mathbf{g} \equiv \mathbf{b}  \Leftrightarrow  \mathbf{D} \mathbf{U}  \mathbf{g} \equiv \mathbf{L^{-1}} \mathbf{b} $.
    Let $\mathbf{h}$ be an $N-1$ by $1$ matrix  such that $\mathbf{h} \equiv \mathbf{U} \mathbf{g}$. 
    Since $\mathbf{U}$ is also an unit, the degree and the number of $\mathbf{g}$ and  $\mathbf{h}$
      are equal~\cite{Butson, Lazebnik}. 
\end{IEEEproof}

In the following, two corollaries of Lemma~\ref{lemma:LDU_like_matrix_solution} are shown.

\begin{corollary}
\label{cor:h_k_e_k_has_solution}
     The linear congruence $\mathbf{D} \mathbf{h} \equiv \mathbf{e}$ has at least one solution, i.e., 
     there exist $h_k$'s such that  $d_{k,k} \cdot h_k \equiv e_k $, where  $1 \leq k \leq N-1$.
\end{corollary}   
\begin{IEEEproof}
    Suppose that for some $k$, there does not exist $h_k$ such that $d_{k,k} \cdot h_k \equiv e_k$.
   Then there does not exist a solution of $\mathbf{D} \mathbf{h} \equiv \mathbf{e}$.
    By Lemma~\ref{lemma:LDU_like_matrix_solution}, there does not exist a  solution of $\mathbf{A} \mathbf{g} \equiv \mathbf{b}$, 
    which contradicts Lemma~\ref{lemma:matrix_equilvalence}. 
\end{IEEEproof}

\begin{corollary}
\label{cor:h_k_e_k_degree_are_equal}
      Let us consider the linear congruence $\mathbf{A} \mathbf{g} \equiv \mathbf{b} $.  There exists a least degree inverse $\mathbf{g}$ such that $\deg\{ \mathbf{g}\} = K$ if and only if $e_{K} \not\equiv 0$ and $e_{k} \equiv 0$, 
      where $K+1 \leq k \leq N-1$. 
\end{corollary}      
\begin{IEEEproof}
\\ ( $\Longrightarrow$  ) \\
Let $\mathbf{g}$ be a least degree inverse such that $\deg\{ \mathbf{g}\} = K$.
 By Lemma~\ref{lemma:LDU_like_matrix_solution}, the degree of  $\mathbf{h}$ is also $K$, i.e., $h_K \not\equiv 0$ and $h_{k} \equiv 0$, where $K+1 \leq k \leq N-1$.
  Since $\mathbf{D} \mathbf{h} \equiv \mathbf{e} $,  $e_{k} \equiv 0$, where $K+1 \leq k \leq N-1$. 
Suppose that $e_{K} \equiv 0$, i.e., $d_{K,K} \cdot h_{K} \equiv 0$. 
 Let us define an $N-1$ by $1$ matrix $\mathbf{h}'$ such that 
  \[ h_{k}' = \left\{ \begin{array}
                 {r@{} l}
                  h_k  &  \mbox{,}\quad     1 \leq  k \leq K-1  \\
                  0    &  \mbox{,}\quad    K \leq k \leq N-1 \\
                 \end{array} \right.   \; . \]  
Then $\mathbf{h}'$   also satisfies the linear congruence $\mathbf{D} \mathbf{h'} \equiv \mathbf{e}$. 
Let $\mathbf{g'}$ be an $N-1$ by $1$ matrix such that $\mathbf{h'} \equiv \mathbf{U} \mathbf{g'}$, 
then $\mathbf{g'}$ is also an inverse.
Since $\deg{\mathbf{\{h'\}}} = \deg{\mathbf{\{g'\}}} < K$  by Lemma~\ref{lemma:LDU_like_matrix_solution}, 
$\mathbf{g}$ cannot be a polynomial of least degree. 
This contradicts the assumption. Consequently, $e_{K} \not\equiv 0$. \\
( $\Longleftarrow$  ) \\
(1)  Suppose that $\deg{\mathbf{\{g\}}} > K$.
Then  by Lemma~\ref{lemma:LDU_like_matrix_solution}, $\deg{\mathbf{\{h\}}} = \deg{\mathbf{\{g\}}} > K$, where  $\mathbf{h} \equiv  \mathbf{U} \mathbf{g} $.
Let us define an $N-1$ by $1$ matrix $\mathbf{h}'$ such that 
  \[ h_{k}' = \left\{ \begin{array}
                 {r@{} l}
                  h_k  &  \mbox{,}\quad     1 \leq  k \leq K  \\
                  0    &  \mbox{,}\quad    K+1 \leq k \leq N-1 \\
                 \end{array} \right.   \; . \]  
Then $\deg{\mathbf{\{h'\}}} = K$ and $\mathbf{h}'$  also satisfies the linear congruence $\mathbf{D} \mathbf{h}' \equiv \mathbf{e}$. 
Then again by Lemma~\ref{lemma:LDU_like_matrix_solution},  $\deg{\mathbf{\{g'\}}} = K$, where $\mathbf{h'} \equiv \mathbf{U} \mathbf{g'}$. 
Since $\deg{\mathbf{\{g'\}}} = K < \deg{\mathbf{\{g\}}} $ and $\mathbf{A} \mathbf{g}' \equiv \mathbf{b}$, $\mathbf{g}$ cannot be a polynomial of least degree. \\
(2) Suppose that $\deg{\mathbf{\{g\}}} < K$.
Since $g_{K} \equiv 0$, $h_{K} \equiv 0$ by Lemma~\ref{lemma:LDU_like_matrix_solution}.
Consequently, $d_{K,K} \cdot h_{K} \equiv e_{K} \equiv 0$. This contradicts the assumption $e_{K} \not\equiv 0$, thus  $\deg{\mathbf{\{g\}}}$ cannot be less than $K$. \\
By (1) and (2), $\deg{\mathbf{\{g\}}} = K$.  
\end{IEEEproof}

In Lemma~\ref{lemma:LDU_like_matrix_solution}, since all the entries of $\mathbf{L}$, $\mathbf{L}^{-1}$, $\mathbf{D}$, $\mathbf{U}$ and $\mathbf{e}$
can be computed for the given $f(x) = f_1x +f_2 x^2$, 
finding the inverse of $f(x)$ reduces to solving $N-1$ linear congruences  $\mathbf{D} \mathbf{h} \equiv \mathbf{e}$ and $\mathbf{h}\equiv  \mathbf{U}  \mathbf{g}$. 
However, the cost of computation for the matrices can be substantial for a large $N$. \\
The computational complexity is shown to be significantly reduced by the following lemma and corollary.
The following lemma shows that the degree of the least degree inverse has an upper bound. 
\begin{lemma}[\cite{JPL:QPP}]
\label{lemma:JPL}
       Let $N = \prod\limits_{p \in \mathcal{P}}  p^{n_{N,p}}$. 
       If $f(x)$ is a QPP, then the inverse PP 
       has degree no larger than $\max\limits_{p \in \mathcal{P}}{n_{N,p}}$.
\end{lemma}      
\begin{IEEEproof}
        Since the set of PPs is a finite group as shown in Lemma~\ref{lemma:matrix_equilvalence},  
        there exists an integer $m$ called an order such that the $m$-fold composition of $f(x)$ with itself is an inverse PP~\cite{D_and_F}.
       Let $f^{(n)}(x)$ be $n$-fold composition of $f(x)$ with itself. 
       It is shown that the coefficient of the degree $k$ term of $f^{(n)}(x)$ is divided by $f_2^{k-1}$ as follows. 
       For $f^{(1)}(x)$, it is clear that $f_2$ divides the coefficient of the degree $2$ term. 
       If the coefficient of the degree $k$ term in $f^{(n)}(x)$ are divisible by $f_2^{k-1}$, then the coefficient of the degree $k$ term in
       $f^{(n+1)}(x) = f_1(f^{(n)}(x))+f_2(f^{(n)}(x))^2$ 
       are also divisible by $f_2^{k-1}$. By induction, the coefficient of the degree $k$ term of $f^{(n)}(x)$ is divided by $f_2^{k-1}$. \\        
       Suppose $k \geq \max\limits_{p \in \mathcal{P}}{n_{N,p}}+1$. Since $f_2$ is divisible by the factors of $N$, 
       $N|f_2^{k-1}$, i.e., $f_2^{k-1} \equiv 0 $. 
       Consequently, there exists an inverse $f^{(n)}(x)$ that contains no terms of degree larger than $\max\limits_{p \in \mathcal{P}}{n_{N,p}}$. 
\end{IEEEproof}
%
%
\begin{corollary}
\label{cor:e_k_is_zero}
     Let us consider the linear congruence $\mathbf{D} \mathbf{h} \equiv \mathbf{e}$ in Lemma~\ref{lemma:LDU_like_matrix_solution}. \\
     For all $k$ such that $k \geq \max\limits_{p \in \mathcal{P}} {n_{N,p}} +1$,  $e_k \equiv 0 $.
\end{corollary}
\begin{IEEEproof}  
   
    Let the degree of the least degree inverse be $K$. 
    Since there exists an inverse such that the degree of the inverse is no larger than $\max\limits_{p \in \mathcal{P}}{n_{N,p}}$ by Lemma~\ref{lemma:JPL},  
    $K \leq \max\limits_{p \in \mathcal{P}}{n_{N,p}}$.
    Consequently,  by Corollary~\ref{cor:h_k_e_k_degree_are_equal},
   $e_k \equiv 0 $, where $K+1 \leq \max\limits_{p \in \mathcal{P}} {n_{N,p}}+1 \leq k \leq N-1 $. 
\end{IEEEproof}
By Corollary~\ref{cor:e_k_is_zero}, only  $\max\limits_{p \in \mathcal{P}} {n_{N,p}}$ by $\max\limits_{p \in \mathcal{P}} {n_{N,p}}$
leading submatrices (the upper-left corners of matrices) of $\mathbf{L}$, $\mathbf{L}^{-1}$, $\mathbf{D}$, $\mathbf{U}$ and  a $\max\limits_{p \in \mathcal{P}} {n_{N,p}}$ by 1 
leading submatrix of $\mathbf{e}$ are required to be computed for finding the inverse of least degree. 
For example, let $N= 2^{18} \cdot 3^{2} \cdot 5$. Then $\max\limits_{p \in \mathcal{P}} {n_{N,p}} = \max \{18,2,1\} = 18$, thus only
$18$ by $18$ leading submatrices of  $\mathbf{L}$, $\mathbf{L}^{-1}$, $\mathbf{D}$, $\mathbf{U}$  
and a $18$ by $1$ leading submatrix of $\mathbf{e}$ need to be computed  instead of 
$N-1$ by $N-1$ submatrices of  $\mathbf{L}$, $\mathbf{L}^{-1}$, $\mathbf{D}$, $\mathbf{U}$  
and a $N-1$ by $1$ submatrix of $\mathbf{e}$.

In the following proposition and corollary, it is shown that the computational complexity for the matrices can be further reduced.  
\begin{proposition}
\label{pro:e_k_value}
      Let $\mathbf{e}$ be an $N-1$ by 1 matrix in Lemma~\ref{lemma:LDU_like_matrix_solution}. 
      Let also $C_{k}$, where $k \geq 0$, be a sequence of integers known as Catalan numbers. 
      The $k$th Catalan numbers are given by 
      \begin{eqnarray}
         C_{k} = \frac{1}{k+1} \binom{2k}{k} = \frac{(2k)!}{(k+1)!  \cdot k!}.  \nonumber
      \end{eqnarray}
      A recurrence relation for $C_{k}$ is 
        \begin{eqnarray}
      C_{k} = \frac{2(2k-1)}{k+1} \cdot C_{k-1},    k \geq 2, \nonumber
        \end{eqnarray}
      i.e., $C_{0}=1, C_{1}=1, C_{2}=2, C_{3}=5, C_{4}=14, C_{5}=42, C_{6}=132.....$
      Then, 
      $$ e_k \equiv  \frac{  k! \cdot C_{k-1} \cdot (-f_2)^{k-1} }{\prod_{m=1}^{2k-1} (f_1+ m f_2)}, $$    
      where $1 \leq k \leq 50$.  
\end{proposition}
\begin{IEEEproof}  
     Let $\mathbf{L}$, $\mathbf{e}$ and $\mathbf{b}$ be the matrices in Lemma~\ref{lemma:LDU_like_matrix_solution}.
     Let also the $k$ by $k$ leading submatrix of  $\mathbf{L}$, $k$ by $1$ leading submatrices of $\mathbf{e}$ and $\mathbf{b}$ be
     $\mathbf{L}'$,  $\mathbf{e}'$ and  $\mathbf{b}'$ respectively.
    
      The following statement, $\mathbf{b}' = \mathbf{L}' \mathbf{e}'$, was verified to be correct for  $1 \leq k \leq 50$.
      \begin{eqnarray} 
       b_k =   k  &= & \sum_{n=1}^{k}  l_{k,n} \cdot e_n \nonumber \\
            &= & \sum_{n=1}^{k} \Big[ \Big\{ \binom{k}{n} \cdot
                       \prod_{m=k}^{k+n-1}(f_1 + m f_2)  \Big\}  \cdot \Big\{
                       \frac{  n! \cdot C_{n-1} \cdot (-f_2)^{n-1}  }{\prod_{m=1}^{2n-1}
                       (f_1+ m f_2)  }   \Big\} \Big]  \nonumber  \\
            &= & \sum_{n=1}^{k}  \Big\{
                       \frac{\prod_{m=k}^{k+n-1}(f_1 + m f_2)}{\prod_{m=1}^{2n-1}
                       (f_1+ m f_2)} \cdot \frac{k!}{(k-n)!} 
                       \cdot  C_{n-1} \cdot (-f_2)^{n-1} \Big\}. \nonumber 
       \end{eqnarray}
       Since $\mathbf{b}' = \mathbf{L}' \mathbf{e}'$, it is clear that $\mathbf{b}' \equiv \mathbf{L}' \mathbf{e}'$. 
       Consequently, $\mathbf{e}' \equiv \mathbf{L'}^{-1} \mathbf{b}'$ for $1 \leq k \leq 50$.
\end{IEEEproof}
\begin{corollary}
\label{cor:e_k_divides_e_k_1}
      Let $N = \prod\limits_{p \in \mathcal{P}}  p^{n_{N,p}} \leq 2^{50}$ 
      and let also $\mathbf{e}$ be an $N-1$ by 1 matrix in Lemma~\ref{lemma:LDU_like_matrix_solution}.
      If $e_k \equiv 0$ for some $k$, then  $e_{n} \equiv 0 $ for $n \geq k+1$. 
\end{corollary}
\begin{IEEEproof}  
 Let $N = \prod\limits_{p \in \mathcal{P}}  p^{n_{N,p}} \leq 2^{50}$, then clearly $\max\limits_{p \in \mathcal{P}}{n_{N,p}} \leq 50 $.
By Lemma~\ref{lemma:JPL} and Proposition~\ref{pro:e_k_value},
\[ e_{k} \equiv \left\{ \begin{array}
                 {r@{} l}
                  \frac{ k! \cdot C(k-1) \cdot (-f_2)^{k-1} }{\prod_{m=1}^{2k-1}
                       (f_1+ m f_2)  }   &  \mbox{,}\quad      1 \leq k \leq 50  \\
                  0    &  \mbox{,}\quad    51 \leq k \leq N-1 \\
                 \end{array} \right.   \; . \]  
         Since $f_1+f_2$ is an unit, $e_1 \equiv \frac{1}{f_1+f_2} \not\equiv 0 $. 
         Suppose that $e_k \equiv 0$ for some $k$.
       Since $C_{k} = \frac{2(2k-1)}{k+1} \cdot C_{k-1}$ for $k \geq 2$, $\frac{(k+1)! \cdot C_{k}}{k! \cdot C_{k-1}} = 2(2k-1)$. 
      Thus $e_{k+1} =  \frac{ 2(2k-1)  \cdot (-f_2) \cdot e_k}{\prod_{m=2k}^{2k+1} (f_1+ m f_2)}   $. 
      Consequently, $e_{k+1} \equiv 0 $ if $e_{k} \equiv 0 $ for some $k$. By induction, 
      if $e_{k} \equiv 0 $ for some $k$,  $e_{n} \equiv 0 $, for $ n \geq k+1$.
\end{IEEEproof}
We are not aware of a closed-form expression of $\mathbf{e}$ when $k$ is larger than $50$. 
However, the investigation on the inverse of a QPP is not restricted under this condition 
since the interleaver size $N$ is far less than $2^{50}$ in practice. 
By Proposition~\ref{pro:e_k_value} and Corollary~\ref{cor:e_k_divides_e_k_1},
matrices $\mathbf{L}$ and $\mathbf{L}^{-1}$ need not to be computed for solving $\mathbf{D} \mathbf{h} \equiv \mathbf{e}$.  
 Combining Lemma~\ref{lemma:matrix_equilvalence},~\ref{lemma:LDU_like_decomposition},
 Proposition~\ref{pro:e_k_value} and Corollary~\ref{cor:e_k_divides_e_k_1} we state the main theorem.  
\begin{theorem}[main Theorem]
\label{thm:main_theorem}
      Let $N = \prod\limits_{p \in \mathcal{P}} p^{n_{N,p}}  \leq 2^{50}$.
      The necessary and sufficient condition for a QPP to admit a least degree inverse $\mathbf{g}$ such that $\deg\{\mathbf{g}\}=K$ is 
      finding a smallest integer $K \geq 1$ such that
      \begin{eqnarray} 
            (K+1)! \cdot C_K \cdot  f_2^{K}  \equiv  0 \mod{N} \nonumber
      \end{eqnarray}

      and the number of inverse PP(s) is 
       \begin{eqnarray} 
            \prod_{k=1}^{K} \gcd(k!, N). \nonumber
      \end{eqnarray}

      Let us slightly abuse the notation in this theorem (and in examples and Algorithm~\ref{tab:algorithm}) 
      by writing $\mathbf{D}$, $\mathbf{U}$, $\mathbf{g}$, $\mathbf{h}$ and $\mathbf{e}$ 
      for $K$ by $K$ leading submatrices $\mathbf{D}$, $\mathbf{U}$ and $K$ by $1$ leading submatrices $\mathbf{g}$, $\mathbf{h}$, $\mathbf{e}$,
      respectively.    
      The inverse PP(s) can be found by using either (1) or (2). \\
      (1) Find all $\mathbf{h}$'s such that $\mathbf{D} \mathbf{h} \equiv \mathbf{e}$ and corresponding $\mathbf{g}$'s such that $\mathbf{h} \equiv \mathbf{U} \mathbf{g}$. \\
      (2) Find a $\mathbf{h}$ such that $h_k \equiv \frac{ C_{k} \cdot (-f_2)^{k-1}    }{\prod_{m=1}^{2k-1} (f_1+ m f_2)  }$, corresponding  $\mathbf{g}$ and add it
      $\prod_{k=1}^{K} \gcd(k!, N)$ zero polynomials. \\     
      Zero polynomials of degree $K$ are 
      $\sum_{k=1}^{K} \big\{ \frac{N}{\gcd{(k!,N)}} \cdot \tau_k \cdot \prod_{m=0}^{k-1}(x-m) \big\}$,
      where $ 0 \leq \tau_k \leq \gcd{(k!,N)}-1$.
 \end{theorem}

\begin{IEEEproof}  
     The necessary and sufficient condition is shown by combining Corollaries~\ref{cor:h_k_e_k_degree_are_equal} and ~\ref{cor:e_k_divides_e_k_1}.
     By Corollary~\ref{cor:h_k_e_k_degree_are_equal}, $\mathbf{g}$ is an inverse of least degree such that $\deg \{\mathbf{g}\} =K$ if and only if $e_{K} \not\equiv 0$ and $e_{k} \equiv 0$ for $K+1 \leq k \leq N-1$.  By Corollary~\ref{cor:e_k_divides_e_k_1},  if $e_k \equiv 0$ for some $k$, then $e_n \equiv 0$ for $ n \geq k+1$. Thus 
    $\mathbf{g}$ is a least degree inverse such that $\deg \{\mathbf{g}\} =K$ if and only if $e_{K} \not\equiv 0$ and $e_{K+1} \equiv 0$. 
    Since $e_1 \not\equiv 0$, finding the degree of the least degree inverse is equivalent to finding the smallest $K$ such that $e_{K+1} \equiv 0$. 
    Consequently, the necessary and sufficient condition\footnote{If $K=1$, $f(x)$ is a linear PP.} for a QPP to admit a least degree inverse is $ (K+1)! \cdot C_K \cdot f_2^{K}  \equiv  0$ since 
    $e_{K+1} \equiv   \frac{ (K+1)! \cdot C_{K} \cdot (-f_2)^{K} }{\prod_{m=1}^{2K+1} (f_1+ m f_2)  }    \equiv 0 \Leftrightarrow    (K+1)! \cdot C_K \cdot f_2^{K}  \equiv  0 $. \\
       The number of solutions of linear congruences $\mathbf{D}  \mathbf{h} \equiv \mathbf{e}$ is  $\prod_{k=1}^{K} \gcd(k!, N)$, since $k$th linear congruence is 
     $d_{k,k} \cdot h_k \equiv e_k$ and $\gcd(d_{k,k},N) = \gcd(k!,N)$. By Lemma~\ref{lemma:LDU_like_matrix_solution},
      the number of solutions of $\mathbf{A} \mathbf{g} \equiv \mathbf{b}$ is also $\prod_{k=1}^{K} \gcd(k!, N)$.  \\
      The complete solution set can be obtained by exhaustively solving  $\mathbf{D} \mathbf{h}  \equiv \mathbf{e}$ and $\mathbf{h} \equiv  \mathbf{U} \mathbf{g}$.
      An alternative is to find one solution $\mathbf{h}$ and $\mathbf{g}$ such that $\mathbf{D} \mathbf{h}  \equiv \mathbf{e}$, $\mathbf{h} \equiv \mathbf{U} \mathbf{g}$
      and add it zero polynomials of degree $K$.  
      Consider $k$th linear congruence $\mathbf{D} \mathbf{h}  \equiv \mathbf{e}$, i.e., $d_{k,k} \cdot h_k \equiv e_k $. Clearly $h_k = \frac{ C_{k}  \cdot (-f_2)^{k-1}    }{\prod_{m=1}^{2k-1} (f_1+ m f_2)  }$
     is a solution of $d_{k,k} \cdot h_k \equiv e_k $, i.e., $k! \cdot h_k \equiv  \frac{ k! \cdot C_{k-1} \cdot (-f_2)^{k-1}  }{\prod_{m=1}^{2k-1} (f_1+ m f_2)} $. 
     The number and form of zero polynomials are shown in Appendix B. 

\end{IEEEproof}
\section{Examples}
%
%
      We present four examples to illustrate the necessary and sufficient conditions of Theorem~\ref{thm:main_theorem}.
      The first and second examples consider interleavers that was investigated in~\cite{Takeshita:AG} and~\cite{LTE:standard}. 
      The third example shows the exact least degree for inverse
      polynomials can be less than an upper bound derived in~\cite{JPL:QPP} and the fourth example shows 
      the necessary and sufficient condition for a QPP  to admit a least degree inverse  $\mathbf{g}$ such that $\deg{\{\mathbf{g\}}} =2$, $3$, $4$ and $5$.\\
      All good quadratic interleavers found in Table~\ref{tab:inverse_for_3gpp_lte}
      admit low degree quadratic inverses. 
      This observation may not be completely surprising
      because ~\cite{Sun, Takeshita:AG} shows that good interleavers should require 
      the second degree coefficient to be relatively large (which works toward satisfying 
      Theorem~\ref{thm:main_theorem}) but bounded by some constraints.
 \\
\begin{enumerate}
      \item  Let $f(x) = f_1x + f_2x^2 \mod{N}$, where 
             $N = 1504 = 2^5 \cdot 47$,  $f_1 =  23$ and $f_2 = 2 \cdot 47 $.
             The smallest $K$ such that $  (K+1)! \cdot C_K \cdot f_2^{K}  \equiv 0 $  is $3$. \\
             By Lemma~\ref{lemma:LDU_like_decomposition}, $3$ by $3$ matrices $\mathbf{D}$, $\mathbf{U}$ and  a $3$ by $1$ matrix $\mathbf{e}$  are computed as follows. 
             \[ \mathbf{D} = 
                \begin{bmatrix} 
                      1   &  0  &  0 \\
                      0   &  2  &  0 \\
                      0   &  0  &  6 \\
                \end{bmatrix},
                \mathbf{U} = 
                \begin{bmatrix} 
                      1   & 117 &  153 \\
                      0   &  1  &  539 \\
                      0   &  0  &   1 \\
                \end{bmatrix},
                \mathbf{e} =
                \begin{bmatrix}
                     797 \\
                     188 \\
                   752  \\
                \end{bmatrix} 
             \]
             %
             Let us now exhaustively solve the equation
             $\mathbf{D} \mathbf{h}  \equiv \mathbf{e}\pmod{1504}$. 
             From $d_{1,1} \cdot h_1 \equiv e_1$, 
             $d_{2,2} \cdot h_2 \equiv e_2$,
             $d_{3,3} \cdot h_3 \equiv e_3$, 
             we can obtain $h_1 = 797$, $h_2 = 94, 846$ and $h_3 = 376, 1128$, respectively. 
             Since $\gcd{(1!,N)} = 1$, $\gcd{(2!,N)} = 2$ and $\gcd{(3!,N)} = 2$, the number of solutions is $4$. 
             Let us choose $h_1 = 797$, $h_2 = 94$ and $h_3 = 376$. 
             We obtain         
             $g_3 = h_3 = 376$, 
             $g_2 = h_2 - u_{23}\cdot g_3 \pmod{1504} = 470$ and
             $g_1 = h_1-u_{12} \cdot g_2 - u_{13} \cdot g_3 \pmod{1504} = 1079$
 by solving $\mathbf{U} \mathbf{g}  \equiv \mathbf{h} \pmod{1504}$. 
             %
             %
             %
       \item  Let $N = 6016= 2^7 \cdot 47$,  $f_1 =  23$ and $f_2 = 2 \cdot 47 $.
              The least degree is $4$. A $4$ by $4$ matrix $\mathbf{U}$ and a $4$ by $1$ matrix $\mathbf{e}$ are computed as follows.  

            \[ 
                \mathbf{U} = 
                \begin{bmatrix} 
                      1   & 117 &  1657  & 1357  \\
                      0   &  1  &  539 & 507     \\
                      0   &  0  &   1  & 1454    \\
                      0   &  0  &   0  &    1    \\
                \end{bmatrix}, 
                 \mathbf{e} = 
                \begin{bmatrix} 
                      3805 \\
                      188  \\
                      752  \\
                      3008 \\
                \end{bmatrix}, 
             \]
             Let us compute $h_k$ such that $h_k = \frac{ C_{k}  \cdot (-f_2)^{k-1}     }{\prod_{m=1}^{2k-1} (f_1+ m f_2)  }$ for each $k$. 
             Then $\mathbf{h} = [ 3805, 94,  4136,  4888]^T$ and  $\mathbf{g} = [  1831, 3854,  1880, 4888]^T$.

      \item  Let $N = 2^{24}$,  $f_1 =  26119$ and $f_2 = 2 \cdot 3 \cdot 41 \cdot 179 $
              The least degree $K$ is $12$, which shows the upper bound $24$ obtained by the technique in~\cite{JPL:QPP} is not tight.\footnote{An inverse $g(x)$ is
     $7612343 x + 4897586 x^2 +  352440 x^3+  2867432 x^4 +  13756448 x^5+
           13890368  x^6 +  915200x^7+  2679424  x^8+   6846976 x^9+    5217280 x^{10}+      53248  x^{11}+   1478656x^{12}.$} 
      \item The necessary and sufficient condition for a QPP to admit a least degree inverse  $\mathbf{g}$ such that $\deg{\mathbf{\{g\}}}=K=2,3,4,5$ is $12f^2_2 \equiv 0$, $120f^3_2 \equiv 0$, $1680f^4_2 \equiv 0$ and $30240f^5_2 \equiv 0$, respectively.  This formula is also shown in~\cite{Suvitie},~\cite{Ryu:inverse_QPP} for $K=2$ and in~\cite{Suvitie} for $K=3$.  \\

\end{enumerate}

188 QPP based interleavers have been proposed in 3GPP LTE~\cite{LTE:standard}. 
Most of the interleavers proposed in~\cite{LTE:standard} admit a quadratic inverse with the exception of $35$
of them.  
In Table~\ref{tab:inverse_for_3gpp_lte}, all of the interleavers that do not admit quadratic inverses are listed 
with their respective inverse PPs of least degree computed using Algorithm~\ref{tab:algorithm}.
\begin{table}[hbt]

   \caption{Algorithm 1}
\label{tab:algorithm}
   \begin{center}
   \begin{small}
   \begin{tabular}{|l|}
   \hline
   \quad\quad An algorithm for finding the inverse PP(s) of least degree  for a QPP $f(x) = f_1 x+ f_2 x^2 \pmod{N}$\\ 
   \hline
   1. If $2|N$, $4 \nmid N$ and $2|f_1$, let $f(x)$ be such that $f(x) = (f_1+\frac{N}{2})x + (f_1+\frac{N}{2})x^2 $. \\
   2. Find the smallest integer $K \geq 1$ such that  $  (K+1)! \cdot C_{K} \cdot  f_2^{K} \equiv 0 $,
       where $C_{0} = 1 $ and $C_{k} = \frac{1}{k+1}\binom{2k}{k}$. \\
       \quad Then, the least degree of the inverse PP(s) is $K$. \\  
   3. Compute $K$ by $K$ matrices $\mathbf{D}$, $\mathbf{U}$ in Lemma~\ref{lemma:LDU_like_decomposition} and  $K$ by $1$ matrix $\mathbf{e}$ 
      in Proposition~\ref{pro:e_k_value}.  \\
   4. There exist two methods for finding the solution set of  $\mathbf{A} \mathbf{g}  \equiv \mathbf{b} \Leftrightarrow \mathbf{D} \mathbf{h}  \equiv \mathbf{e}, \mathbf{h} \equiv \mathbf{U}  \mathbf{g}         $. \\
    \quad (1) All the $ \mathbf{h}$'s and $ \mathbf{g}$'s can be found by solving $K$ linear congruences $\mathbf{D} \mathbf{h} \equiv \mathbf{e} $ and 
                 $ \mathbf{h} \equiv \mathbf{U} \mathbf{g} $. \\
      \quad  \quad  \quad       $ \mathbf{g}$'s  can be computed by by back-substitution. \\
        \quad  \quad  \quad      Note that $g_K = h_K$ and $g_k =  h_k - \sum_{m=k+1}^{K} u_{k,m} \cdot g_{m} $ for $1 \leq k \leq K-1$. \\
         %
   \quad  (2) Find one inverse and add it zero polynomials of degree $K$. \\
    \quad \quad \quad   Compute $h_k =  \frac{ C_k \cdot (-f_2)^{k-1}  }{\prod_{m=1}^{2k-1} (f_1+ m f_2)  } $ for $1 \leq k \leq K$ and corresponding $\mathbf{g}$ such that 
                     $\mathbf{h} \equiv \mathbf{U} \mathbf{g} $.          \\
     \quad \quad \quad     Convert $K$ by $1$ matrix $\mathbf{g}$ into a polynomial and add it
           $z(x) = \sum_{k=1}^{K} \big\{ \frac{N}{\gcd{(k!,N)}} \cdot \tau_k \cdot \prod_{m=0}^{k-1}(x-m) \big\} $,\\
         \quad \quad \quad       where $1 \leq \tau_k \leq \gcd(k!, N)-1$.\\
   \hline
   \end{tabular}
   \end{small}
   \end{center}
\end{table}
\begin{table}[!t]
\caption{Inverse PPs of Least Degree for 3GPP LTE Interleavers without Quadratic Inverses}
\label{tab:inverse_for_3gpp_lte}
\centering
\begin{tabular}{|c|c|c||c|c|c|   }
\hline
length  & QPP & An Inverse PP of Least Degree & length  & QPP & An Inverse PP of Least Degree\\
\hline 
 928  &  $15x+58x^2$     &   $31x+   290x^2+  232x^3$    &4544 &  $357x+142x^2$    &   $4509x+  994x^2+   2840x^3$ \\
 1056 &  $17x+66x^2$     &   $1025x+ 726x^2+  792x^3$    &4672 &  $37x+146x^2$     &   $2557x+  1022x^2+  4088x^3$\\
 1184 &  $19x+74x^2$     &   $779x+  74x^2+   296x^3$    &4736 &  $71x+444x^2$     &   $2935x+  3996x^2+  3552x^3$\\
 1248 &  $19x+78x^2$     &   $427x+  78x^2+   936x^3$    &4928 &  $39x+462x^2$     &   $1927x+  1078x^2+  616x^3$  \\
 1312 &  $21x+82x^2$     &   $781x+  574x^2+  984x^3$    &4992 &  $127x+234x^2$    &   $511x+   2730x^2+  4056x^3+  2184x^4$\\
 1376 &  $21x+86x^2$     &   $557x+  602x^2+  344x^3$    &5056 &  $39x+158x^2$     &   $3079x+  2054x^2+  1896x^3$\\
 1504 &  $49x+846x^2$    &   $353x+  282x^2+  376x^3$    &5184 &  $31x+96x^2$      &   $3679x+  1632x^2+  1152x^3$\\ 
 1632 &  $25x+102x^2$    &   $1273x+ 306x^2+  408x^3$    &5248 &  $113x+902x^2$    &   $2833x+  410x^2+   4264x^3+  328x^4$\\
 1696 &  $55x+954x^2$    &   $663x+  530x^2+  424x^3$    &5312 &  $41x+166x^2$     &   $3401x+  498x^2+   3320x^3        $ \\
 1760 &  $27x+110x^2$    &   $163x+  990x^2+  1320x^3$   &5440 &  $43x+170x^2$     &   $1107x+  1530x^2+  680x^3       $\\
 1824 &  $29x+114x^2$    &   $1541x+ 1710x^2+ 1368x^3$   &5504 &  $21x+86x^2$      &   $2621x+  5074x^2+  1032x^3+  5160x^4$\\
 1888 &  $45x+354x^2$    &   $21x+   1534x^2+ 472x^3$    &5568 &  $43x+174x^2$     &   $1651x+  1566x^2+  4872x^3         $\\
 1952 &  $59x+610x^2$    &   $579x+  1586x^2+ 488x^3$    &5696 &  $45x+178x^2$     &   $3829x+  5518x^2+  4984x^3        $\\
 2112 &  $17x+66x^2$     &   $1025x+ 1782x^2+ 792x^3$    &5824 &  $89x+182x^2$     &   $409x+   3458x^2+  3640x^3       $\\
 2944 &  $45x+92x^2$     &   $1701x+ 1748x^2+ 2208x^3$   &5952 &  $47x+186x^2$     &   $95x+    930x^2+   3720x^3      $\\ 
 4160 &  $33x+130x^2$    &   $3057x+ 1430x^2+ 1560x^3$   &6016 &  $23x+94x^2$      &   $1831x+  3854x^2+  1880x^3+  4888x^4$ \\ 
 4288 &  $33x+134x^2$    &   $3281x+ 1474x^2+ 2680x^3$   &6080 &  $47x+190x^2$     &   $2943x+  950x^2+   2280x^3         $  \\
 4416 &  $35x+138x^2$    &   $347x+  2346x^2+ 552x^3$    &     &                   &                        \\
\hline
\end{tabular}
\end{table}
\section{Conclusion}
We derived in Theorem~\ref{thm:main_theorem} a necessary and sufficient
condition to determine the least degree inverse for a
QPP. We also provided an algorithm to explicitly compute the inverse PP(s). \\
188 QPP interleavers were proposed in 3GPP LTE~\cite{LTE:standard}. Most of the QPP interleavers in~\cite{LTE:standard} admit a QPP inverse. We applied the theory in this correspondence to tabulate all inverse PPs of degree larger than two. 
Further, it was shown that inverses of good interleavers in~\cite{LTE:standard} have low degrees and a possible explanation is given. 
\newpage
\appendix
(A) [Lemma~\ref{lemma:LDU_like_decomposition}] \\
     We use two-fold induction and prove $\mathbf{A}= \mathbf{L}  \mathbf{D} \mathbf{U} $ by showing that column-reduced form of $\mathbf{A}$ is equivalent to  $ \mathbf{L} \mathbf{D} $. \\ 
        Let us define an $N-1$ by $N-1$ elementary matrix $\mathbf{T}^{(i,j)}$ such that
            \[ t^{(i,j)}_{m,n} = \left\{ \begin{array}
                 {r@{\quad \quad}l}
                  1       & \mbox{if  }       m = n  \\
              -u_{i,j}    & \mbox{if  }       m=i, n=j \\
                  0       & \mbox{otherwise}
                                 \end{array} \right.,   \;  \] 
     where $1 \leq i \leq j-1$, $2 \leq j \leq N-1$ and $1 \leq m,n \leq N-1$. \\
     Let
     $\mathbf{T} = \mathbf{T}^{(1,2)} \cdot  \mathbf{T}^{(1,3)}   \cdot  \mathbf{T}^{(2,3)} \cdots   \mathbf{T}^{(1,N-1)}  \cdots   \mathbf{T}^{(N-2,N-1)}  $,  
     then it is easily verified that $\mathbf{T} = \mathbf{U}^{-1}$. \\
     Let us also define $N-1$ by $N-1$ lower triangular matrices $\mathbf{L}^{(i,j)}$ such that 
  $$\mathbf{L}^{(i,j)} = \mathbf{A }\mathbf{T}^{(1,2)} \cdot  \mathbf{T}^{(1,3)} \cdot  \mathbf{T}^{(2,3)} \cdots   \mathbf{T}^{(1,j)}  \cdots   \mathbf{T}^{(i-1,j)}    \mathbf{T}^{(i,j)}, \mbox{i.e.,}  $$
\begin{equation}
\mathbf{L}^{(i,j)} = \begin{cases}
                        \mathbf{L}^{(j-1,j)} \mathbf{T}^{(1,j)}  & \text{if   }   i = 1 \nonumber \\
                        \mathbf{L}^{(i-1,j)} \mathbf{T}^{(i,j)}   &   \text{if   }  2 \leq i  \leq j-1  \nonumber   
              \end{cases}
\end{equation}
   
  Since  $\mathbf{U}$ is an unit,  $\mathbf{A} = \mathbf{L} \mathbf{D} \mathbf{U} $ if and only if 
  \begin{eqnarray}
  \mathbf{A} \mathbf{U}^{-1} = \mathbf{A} \mathbf{T} & =& \underbrace{ \mathbf{A} \mathbf{T}^{(1,2)} \cdot  \mathbf{T}^{(1,3)}   \cdot  \mathbf{T}^{(2,3)} \cdots  \mathbf{T}^{(i,j)}}_{\mathbf{L}^{(i,j)}}  \cdots  \mathbf{T}^{(1,N-1)}  \cdots   \mathbf{T}^{(N-2,N-1)} \nonumber \\
    & = & \mathbf{L}^{(N-2,N-1)} =   \mathbf{L} \mathbf{D}. \label{AU_LD}      
  \end{eqnarray}

We use induction on $j$ and prove eq.~\eqref{AU_LD} by showing that  $\mathbf{L}^{(j-1,j)}$ is as follows.
\begin{equation}
\label{eq:induction_on_j}
l_{m,n}^{(j-1,j)} = \begin{cases}
                       n! \cdot l_{m,n}  & \text{if   }   1 \leq n \leq j  \\
                              a_{m,n}    &   \text{if   } j+1 \leq n \leq N-1    
              \end{cases}.
\end{equation}
Upon completion of column reduction, $j=N-1$, thus eq.~\eqref{AU_LD} holds. \\
We first show that eq.~\eqref{eq:induction_on_j} holds for $j=2$. \\
  By definition, $\mathbf{L}^{(1,2)} = \mathbf{A} \mathbf{T}^{(1,2)} $. Since $t^{(1,2)}_{1,2} = -u_{1,2} $ and 
      $u_{1,2}   =  \mathbf{q}^{(1,2)} \mathbf{V}^{(1,2)} \mathbf{r}^{(2)} =f_1 +f_2  $,
      \begin{eqnarray}
            l^{(1,2)}_{m,2} & = &   -u_{1,2} \cdot a_{m,1} + a_{m,2}                                 \nonumber \\ 
                       & = &    - (f_1+ f_2) \cdot (m f_1+ m^2 f_2) + (m f_1+ m^2 f_2)^2  \nonumber \\
                         & = &     (m f_1+ m^2 f_2) \cdot \{(m-1) f_1+ (m-1)(m+1) f_2\}  \nonumber \\
                       & = &  (m-1) \cdot m \cdot ( f_1+ f_2)\cdot \{ f_1 +(m+1)f_2  \} \nonumber 
      \end{eqnarray}
  Consequently, 
\begin{equation}
l_{m,2}^{(1,2)} = \begin{cases}
                          0   & \text{if   }   m = 1  \nonumber  \\
                         2!\cdot  l_{m,2}    &   \text{if   } m \geq 2   \nonumber 
              \end{cases}
\end{equation}
Thus eq.~\eqref{eq:induction_on_j} holds for $j=2$.
Suppose now that eq.~\eqref{eq:induction_on_j} holds for $j \geq 2$.
For each $j$, we use induction on $i$ and show that eq.~\eqref{eq:induction_on_i} holds. 
Upon completion of induction on $i$, we show eq.~\eqref{eq:induction_on_j} holds for $j+1$. 
\begin{eqnarray}
            l^{(i,j+1)}_{m,j+1} = \big[\prod_{k=0}^{i}(m-k)\{f_1+(m+k)f_2\}\big] \cdot  \mathbf{q}^{(m,j+1)} \mathbf{V}^{(i+1,j+1)} \mathbf{r}^{(j+1)}   \label{eq:induction_on_i}
      \end{eqnarray}

   In the following, $\mathbf{L}^{(i,j+1)}$  is  shown below in matrix form. 
$$  
  \bordermatrix{
            & col_1                & col_2                & \cdots   & col_{j}             & col_{j+1}                & col_{j+2}   & \cdots   & col_{N-1}   \cr
  low_1     & 1! \cdot l_{1,1}     & 0                    & \ldots   & 0                   &  0                       & a_{1,j+2}   & \ldots   & a_{1,N-1}  \cr  
  low_2     & 1! \cdot l_{2,1}     & 2! \cdot l_{2,2}     & \ldots   & 0                   &  0                       & a_{2,j+2}   & \ldots   & a_{2,N-1}   \cr  
  \vdots    & \vdots               & \vdots               & \ddots   & \vdots              & \vdots                   & \vdots      & \ddots   & \vdots   \cr  
  low_i     & 1! \cdot l_{i,1}     & 2! \cdot l_{i,2}     & \ldots   & 0                   & l^{(i,j+1)}_{i,j+1}=0    & a_{i,j+2}   & \ldots   & a_{i,N-1}   \cr  
  low_{i+1} & 1! \cdot l_{{i+1},1} & 2! \cdot l_{{i+1},2} & \ldots   & 0                   & l^{(i,j+1)}_{i+1,j+1}    & a_{i+1,j+2} & \ldots   & a_{i+1,N-1} \cr 
  \vdots    & \vdots               & \vdots               & \ddots   & \vdots              & \vdots                   & \vdots      & \vdots   & \vdots  \cr 
  low_{j}   & 1! \cdot l_{j,1}     & 2! \cdot l_{j,2}     & \ldots   & j! \cdot l_{j,j}    & l^{(i,j+1)}_{j,j+1}      & a_{j,j+2}   & \ldots   & a_{j,N-1} \cr 
  \vdots    & \vdots               & \vdots               & \ddots   & \vdots              & \vdots                   & \vdots      & \ddots   & \vdots   \cr
  low_{N-1} & 1! \cdot l_{N-1,1}   & 2! \cdot l_{N-1,2}   & \ldots   & j! \cdot l_{N-1,j}  & l^{(i,j+1)}_{N-1,j+1}    & a_{N-1,j+2} & \ldots   & a_{N-1,N-1} \cr 
  }
$$
The elementary matrix $\mathbf{T}^{(1,j+1)}$ subtracts $u_{1,j+1}$ times column $1$ from column $j+1$ of  $\mathbf{L}^{(j-1,j)}$. 
  We show that  $\mathbf{L}^{(j-1,j)}$ multiplied by $\mathbf{T}^{(1,j+1)}$ leaves other columns unchanged except the column $j+1$ and creates a zero in the $(1,j+1)$  position of $\mathbf{L}^{(1,j+1)} = \mathbf{L}^{(j-1,j)} \mathbf{T}^{(1,j+1)}$.  \\
  When $i=1$, eq.~\eqref{eq:induction_on_i} holds, since 
      \begin{eqnarray}
            l^{(1,j+1)}_{m,j+1}   & = &   -u_{1,j+1} \cdot l^{(j-1,j)}_{m,1} + l^{(j-1,j)}_{m,j+1}                                 \nonumber \\ 
                          & = &   -u_{1,j+1} \cdot 1! \cdot l_{m,1} + a_{m,j+1}                                 \nonumber \\ 
                          & = & - \mathbf{q}^{(1,j+1)} \mathbf{V}^{(1,j+1)} \mathbf{r}^{(j+1)}  \cdot \binom{m}{1}(f_1+mf_2 )+ (mf_1+m^2f_2)^{j+1}   \nonumber \\
                          & = &  - (f_1+f_2)^{j} \cdot (mf_1+m^2f_2) +  (mf_1+m^2f_2)^{j+1}    \nonumber \\
                          & = &   (mf_1+m^2f_2) \cdot \big\{  (mf_1+m^2f_2)^{j} -  (f_1+f_2)^{j}     \big\}       \nonumber \\
                          & = &   (mf_1+m^2f_2) \cdot \big\{ (m-1)f_1 + (m-1)(m+1)f_2  \big\} \cdot \sum^{j-1}_{k=0} \big\{ (mf_1+m^2f_2)^{j-1-k} \cdot (f_1+f_2)^{k}  \big\}  \nonumber \\
                          & = &   m(m-1)(f_1+mf_2)\{f_1+(m+1)f_2 \} \cdot \mathbf{q}^{(m,j+1)}\mathbf{W}^{(1,j+1)}\mathbf{r}^{(j+1)}         \nonumber \\
                          & = &   \big[\prod_{k=0}^{1}(m-k)\{f_1+(m+k)f_2\}\big] \cdot  \mathbf{q}^{(m,j+1)} \mathbf{V}^{(2,j+1)} \mathbf{r}^{(j+1)}.   \nonumber 
      \end{eqnarray}
     %
     \noindent Thus $l^{(1,j+1)}_{1,j+1} =0$ as desired.  \\   
      Suppose now that eq.~\eqref{eq:induction_on_i} holds for $i$.  $\mathbf{T}^{(i+1,j+1)}$  subtracts $u_{i+1,j+1}$ times column $i+1$ from column $j+1$ of  $\mathbf{L}^{(i,j+1)}$, where $2 \leq i \leq j$.
  In the following, it is shown that that $\mathbf{L}^{(i,j+1)}$ multiplied by $\mathbf{T}^{(i+1,j+1)}$ leaves other columns unchanged except the column $j+1$ and creates a zero in the $(i+1,j+1)$ position of $\mathbf{L}^{(i+1,j+1)} = \mathbf{L}^{(i,j+1)} \mathbf{T}^{(i+1,j+1)}$.  \\
      %
      \begin{eqnarray}
      l^{(i+1,j+1)}_{m,j+1} &= & -u_{i+1,j+1} \cdot  l^{(i,j+1)}_{m,i+1}  + l^{(i,j+1)}_{m,j+1}    \nonumber \\
            &=& -u_{i+1,j+1} \cdot  (i+1)! \cdot l_{m,i+1} + l^{(i,j+1)}_{m,j+1}    \nonumber \\
            &=& -\mathbf{q}^{(i+1,j+1)} \mathbf{V}^{(i+1,j+1)} \mathbf{r}^{(j+1)} \cdot (i+1)! \cdot \binom{m}{i+1} \cdot   \prod_{k=m}^{m+i} (f_1+k f_2)  +  \nonumber \\
            & &    \big[\prod_{k=0}^{i}(m-k)\{f_1+(m+k)f_2\} \big] \cdot \mathbf{q}^{(m,j+1)} \mathbf{V}^{(i+1,j+1)} \mathbf{r}^{(j+1)}   \nonumber \\
            &=&  -\mathbf{q}^{(i+1,j+1)} \mathbf{V}^{(i+1,j+1)} \mathbf{r}^{(j+1)}  \cdot \big[\prod_{k=0}^{i}(m-k)\{f_1+(m+k)f_2\}\big]  +   \nonumber \\
            & &       \big[\prod_{k=0}^{i}(m-k)\{f_1+(m+k)f_2\}\big] \cdot \mathbf{q}^{(m,j+1)} \mathbf{V}^{(i+1,j+1)} \mathbf{r}^{(j+1)} \nonumber \\
            &=& \big[\prod_{k=0}^{i}(m-k)\{f_1+(m+k)f_2\}\big] \cdot \{\mathbf{q}^{(m,j+1)} - \mathbf{q}^{(i+1,j+1)} \}  \mathbf{V}^{(i+1,j+1)} \mathbf{r}^{(j+1)}. \nonumber 
      \end{eqnarray}
      \begin{eqnarray}
            & &\mathbf{q}^{(m,j+1)} -  \mathbf{q}^{(i+1,j+1)} \nonumber \\
            &=& \big[1, mf_1+m^2f_2,...,(mf_1+m^2f_2)^{j}  \big] -   \big[1, (i+1)f_1+(i+1)^2f_2,...,\{(i+1)f_1+(i+1)^2f_2\}^{j}   \big]  \nonumber \\
            &=&   \{m-(i+1)\}\{f_1+(m+i+1)f_2\}  \nonumber \cdot \\
            & & \big[0, 1,  m(f_1+mf_2) + (i+1)\{f_1+(i+1)f_2\}, ...,  \sum_{n=0}^{j-1} \{m(f_1+mf_2)\}^{j-1-n}\{(i+1)(f_1+(i+1)f_2)\}^{n} \big] \nonumber \\
            &=&    \{m-(i+1)\}\{f_1+(m+i+1)f_2\} \cdot \mathbf{q}^{(m,j+1)}   \mathbf{W}^{(i+1,j+1)}.   \nonumber
      \end{eqnarray}
      Thus, 
      \begin{eqnarray}
                         &    & l^{(i+1,j+1)}_{m,j+1}    \nonumber \\            
                           & = & \big[\prod_{k=0}^{i}(m-k)\{f_1+(m+k)f_2\}\big] \cdot    \{m-(i+1)\}\{f_1+(m+i+1)f_2\} \cdot \nonumber  \\
                          &  &   \mathbf{q}^{(m,j+1)} \mathbf{W}^{(i+1,j+1)}  \mathbf{V}^{(i+1,j+1)}    \mathbf{r}^{(j+1)} \nonumber \\
                         & = & \prod_{k=0}^{i+1}(m-k)\{f_1+(m+k)f_2\} \cdot \mathbf{q}^{(m,j+1)} \mathbf{V}^{(i+2,j+1)} \mathbf{r}^{(j+1)}  \nonumber 
      \end{eqnarray}
      Consequently, $l^{(i+1,j+1)}_{i+1,j+1} = 0$ and eq.~\eqref{eq:induction_on_i} holds for $i+1$.\\
      We now show that eq.~\eqref{eq:induction_on_j} holds for $j+1$.  
      Let $i=j$ in eq.~\eqref{eq:induction_on_i}. Then
 \begin{eqnarray}
            & & \mathbf{q}^{(m,j+1)} \mathbf{V}^{(j+1,j+1)} \nonumber \\
            &=& \big[1, mf_1+m^2f_2,...,(mf_1+m^2f_2)^{j}  \big] \cdot \mathbf{W}^{(j,j+1)} \mathbf{W}^{(j-1,j+1)} \mathbf{W}^{(j-2,j+1)} \cdots \mathbf{W}^{(1,j+1)}  \nonumber \\
            &=& \big[0, 1, ... ] \cdot  \mathbf{W}^{(j-1,j+1)} \mathbf{W}^{(j-2,j+1)} \cdots \mathbf{W}^{(1,j+1)}         \nonumber \cdot \\
            &= &  \big[0, 0, 1, ... ] \cdot   \mathbf{W}^{(j-2,j+1)} \cdots \mathbf{W}^{(1,j+1)} \nonumber \\
            &  &  \cdots  \nonumber \\
            &=&   [0, 0,...,1].   \nonumber
      \end{eqnarray}     
    Thus $\mathbf{q}^{(m,j+1)} \mathbf{V}^{(j+1,j+1)}  \mathbf{r}^{(j+1)}=1 $. 
     Consequently, $ l^{(j,j+1)}_{m,j+1} = 0$, where  $m \leq j$ and 
    \begin{eqnarray}
               l^{(j,j+1)}_{m,j+1}    &  =  & \prod_{k=0}^{j}(m-k)\{f_1+(m+k)f_2\}  \nonumber \\            
           & = & (j+1)! \cdot \binom{m}{j+1} \cdot \prod_{k=m}^{m+j}(f_1+k f_2) \nonumber  \\
                          & = &   (j+1)! \cdot l_{m,j+1}, \nonumber
      \end{eqnarray}
     where $j+1 \leq m \leq N-1$. Consequently eq.~\eqref{eq:induction_on_j} holds for $j+1$. \\           
(B) [The number and the form of zero polynomials of degree $K$] \\
 We show the number and the explicit form of zero polynomials of degree $K$, where $K \leq N-1$.\footnote{A different proof is shown in~\cite[pp. 245]{Kempner} and~\cite{Lee:null_poly} for the explicit form of zero polynomials.} \\
In Lemma~\ref{lemma:TFAE}, the necessary and sufficient conditions for a polynomial to be a zero polynomial is shown and 
in Lemma~\ref{lemma:number_of_zero_polynomial} and~\ref{lemma:form_of_zero_polynomial}, the number and the explicit form of zero polynomials of degree $K$ are derived by using Lemma~\ref{lemma:TFAE}.  \\
Let us define $z_n(x)$, where $0 \leq x \leq N-1$ as follows.

  \[ z_n(x) \equiv \left\{ \begin{array}
                 {r@{\quad \quad}l}
                  z(x) =  \sum_{k=1}^{K} z_k x^k       & \mbox{if  }       n = 0  \\
             z_{n-1}(x+1\bmod{N}) - z_{n-1}(x\bmod{N})   & \mbox{if  }   1 \leq n \leq K
                                 \end{array} \right. .   \; 
  \] 
  
       \begin{lemma} \label{lemma:zero_poly_condition}
        \label{lemma:TFAE}      
        The following statements are equivalent. \\
        (1) $z(x) \equiv 0, \mbox{where } 0 \leq x \leq N-1 $.  \label{lemma:TFAE1} \\
        (2) $z(x) \equiv 0,  \mbox{where } 0 \leq x \leq K $.  \label{lemma:TFAE2} \\
        (3) $z_n(0) \equiv 0,  \mbox{where } 0 \leq n \leq K$.        \label{lemma:TFAE3} 
      \end{lemma}
\begin{IEEEproof}\\
 ( (1) $\Longrightarrow$ (2) ) \\
Trivial. \\
 ( (2) $\Longrightarrow$ (3) ) \\
It is easily shown by induction that if $z_n(x) \equiv 0$,  where $0 \leq x \leq K-n  $,
then $z_{n+1}(0) \equiv 0$, where $0 \leq x \leq K- (n+1)$. Consequently, (3) holds.  \\
 ( (3) $\Longrightarrow$ (1) ) \\
Suppose that  $z_n(0) \equiv 0 $, where $0 \leq n \leq K$. 
Since $z_K(x)$ is a constant, if $z_K(0) \equiv 0$, then $z_K(x) \equiv 0 $, where  $ 1 \leq  x \leq N-1$. \\
Consider $z_{K}(x) \equiv z_{K-1}(x+1) - z_{K-1}(x)$. Since $z_{K-1}(0) \equiv 0$ and $z_{K}(0) \equiv 0$ by assumption, $z_{K-1}(1) \equiv z_{K}(0) + z_{K-1}(0) \equiv 0$. Then by induction on $x$, it is shown that $z_{K-1}(x) \equiv 0$ for  $ 1 \leq  x \leq N-1$. \\ 
The induction outlined above are then repeated for $n=K-2,K-3,...,2,1$. Hence, (1) holds as desired.
\end{IEEEproof}
  \begin{lemma} 
      \label{lemma:number_of_zero_polynomial}
       The number of zero polynomials of degree $K$ is $\prod_{k=1}^{K} \gcd(k!, N)$. 
  \end{lemma}
\begin{IEEEproof}\\
       Let $\mathbf{\bar{A}}$  be a $K$ by $K$ leading submatrix of $\mathbf{A}$ in Lemma~\ref{lemma:LDU_like_decomposition} and let $f_1$, $f_2$ be $1$ and $0$ respectively. \\ 
Let also $\mathbf{\bar{L}}$, $\mathbf{\bar{D}}$, $\mathbf{\bar{U}}$, $\mathbf{\bar{h}}$ and $\mathbf{z}$ be the corresponding leading submatrices of  $\mathbf{L}$, $\mathbf{D}$, $\mathbf{U}$,  $\mathbf{h}$ and a zero polynomial of degree $K$.
  Then $z(x) \equiv 0$, where $0 \leq x \leq K$ is equivalent to 
$ \mathbf{\bar{A}}\mathbf{z} \equiv \mathbf{0}$, where $\bar{a}_{i,j} = i^j \pmod{N}$, 
$\mathbf{z} =  [z_1,z_2,...,z_K]^T \equiv \mathbf{\bar{U}^{-1}}\mathbf{\bar{h}}$ and $\mathbf{0}$ is a $K$ by $1$ zero matrix. 
This is shown by evaluating $z(x) \equiv 0$ at each point $1 \leq x \leq N-1$.\\
In Lemma~\ref{lemma:LDU_like_matrix_solution}, it is shown that $\mathbf{L}$ and $\mathbf{U}$ are units. It also holds for $\mathbf{\bar{L}}$ and $\mathbf{\bar{U}}$ since all the elements of  $\mathbf{\bar{L}}$ and $\mathbf{\bar{U}}$ on the diagonal are $1$s. 
Then by Theorem~\ref{thm:linear_congruence} and Lemma~\ref{lemma:LDU_like_matrix_solution}, the number of zero polynomials of degree $K$ is the number of solutions of $\mathbf{\bar{D}}\mathbf{\bar{h}} \equiv \mathbf{0}$, i.e., $\prod_{k=1}^{K} \gcd(k!, N)$. \\
The set of solutions of $\mathbf{A} \mathbf{g} \equiv \mathbf{b}$ ($\Leftrightarrow  \mathbf{A} \mathbf{g}+\mathbf{z} \equiv \mathbf{b}$), where $\mathbf{g}$ has degree $K$, is therefore composed of one particular solution $\mathbf{g}$ and zero polynomials of degree $K$.

%
\end{IEEEproof}

\begin{lemma}
      \label{lemma:form_of_zero_polynomial}
       Zero polynomials of degree $K$ are of the form 
       \begin{eqnarray}
      z(x) =  \sum_{k=1}^{K} \big\{ \frac{N}{\gcd{(k!,N)}} \cdot \tau_k \cdot \prod_{m=0}^{k-1}(x-m) \big\}, \mbox{ where }   0 \leq \tau_k \leq  \gcd{(k!,N)}-1.  \nonumber 
       \end{eqnarray}    
  \end{lemma}
\begin{IEEEproof}\\
Consider  $K$th linear congruence of $\mathbf{\bar{D}}\mathbf{\bar{h}} \equiv \mathbf{0}$ in Lemma~\ref{lemma:number_of_zero_polynomial}, i.e., $\bar{d}_{K,K} \cdot \bar{h}_K \equiv 0$.
Since $\bar{d}_{K,K} = K!$, by Theorem~\ref{thm:linear_congruence} and Lemma~\ref{lemma:LDU_like_matrix_solution}, $\bar{h}_K = z_K = \frac{N}{\gcd(K!,N)} \cdot \tau_K$, where $0 \leq \tau_K \leq \gcd(K!,N)-1$.
Suppose now that $z^{(K)}(x) $ is a zero polynomial of degree $K$, then $z^{(K)}(x) = \frac{N}{\gcd{(K!,N)}} \cdot \tau_K \cdot z'(x) $, where $z'(x)$ is a monic polynomial of degree $K$. \\
 Let $z'(x) = \prod_{m=0}^{K-1}(x-m)$ and consider 
$z^{(K)}(x) = \frac{N}{\gcd{(K!,N)}} \cdot \tau_K \cdot \prod_{m=0}^{K-1}(x-m)$. 
It is clear that $z^{(K)}(x) \equiv 0$, where $0 \leq x \leq K-1$.
 Further $z^{(K)}(K) = \frac{N}{\gcd{(K!,N)}}\cdot \tau_K \cdot K! = N \cdot \tau_K \cdot \frac{K!}{\gcd{(K!,N)}} \equiv 0$, thus $z^{(K)}(x) \equiv 0$, where $0 \leq x \leq K$. 
Consequently, by eq. (2) in Lemma~\ref{lemma:TFAE}, $z^{(K)}(x)$ is a zero polynomial of degree $K$. 
Since $\tau_m \neq \tau_n$, where $m \neq n$, $z^{(K)}(x)$'s are equivalent but not congruent polynomials. \\
Let us now consider ($K-1$)th and $K$th linear congruences of $\mathbf{\bar{D}}\mathbf{\bar{h}} \equiv \mathbf{0}$,
where  $\bar{h}_{K-1,K-1} \equiv z_{K-1} \equiv \frac{N}{\gcd((K-1)!,N)} \cdot n_{K-1}$ and $\bar{h}_{K,K} \equiv 0$.
By using a similar argument above, it is shown that $z^{(K-1)}(x) = \frac{N}{\gcd{((K-1)!,N)}} \cdot \tau_{K-1} \cdot \prod_{m=0}^{K-2}(x-m)$ is a zero polynomial of degree $K-1$. Note that $z^{(K)}(x) + z^{(K-1)}(x)$ is also a zero polynomial of degree $K$.\\
Applying this repeatedly for $k=K-2,K-3,...,2$, the desired result follows, i.e., zero polynomials of degree $K$ are of the form  $\sum_{k=1}^{K} z^{(k)}(x) =  \sum_{k=1}^{K} \big\{ \frac{N}{\gcd{(k!,N)}} \cdot \tau_k \cdot \prod_{m=0}^{k-1}(x-m) \big\}$.\footnote{It is easily verified that there does not exist a non-trivial zero polynomial of degree $1$}     
\end{IEEEproof}

\end{document}